\documentclass[a4paper]{article}
\usepackage{graphicx} % Used for displaying a sample figure. If possible, figure files should be included in EPS format.
\usepackage{xcolor}
\usepackage{comment}
\usepackage[T1]{fontenc}

\usepackage{xcolor}
\usepackage{geometry} % Smaller margins 1/2
\usepackage{hyperref} % hyperlinks
\hypersetup{colorlinks,allcolors=black} % Remove green box from hyperlinks
\usepackage{caption} % captions
\usepackage{subcaption} % subcaptions
\usepackage[style=nature]{biblatex} %Imports biblatex package, sorted by order of appearance
\usepackage{amsmath} % math formatting
\usepackage{amsfonts} % formatting for mathbb
\usepackage{amssymb} % more math symbols
\usepackage[affil-it]{authblk} %adds affiliation
\usepackage{titlesec} % package for setting section spacing
\titlespacing*{\section}{0pt}{1em}{0em} % section spacing {left}{before}{after}
\titlespacing*{\subsection}{0pt}{1em}{0em} % subsection spacing {left}{before}{after}
\usepackage[title,titletoc]{appendix} %add "Appendix" before section letter (like "Appendix A" instead of just "A") in both the appendix section headings and the TOC

\usepackage{tocloft}
\usepackage{braket}

\usepackage{float}
\usepackage{todonotes}
\title{Finite path integrals on stochastic branched structures}

\author[1]{Roukaya Dekhil%
    \thanks{\texttt{roukaya.dekhil@unifi.it}}
    }
\author[2]{Clifford Ellgen%
    \thanks{\texttt{cellgen@alumni.caltech.edu}}
}
\author[3]{Bruno Klajn%
    \thanks{\texttt{bruno.klajn@pbf.unizg.hr}, Corresponding author}
    }
    
\affil[1]{Universit\`a degli Studi di Firenze, Piazza di San Marco, 4, \newline 50121 Firenze FI, Italy, EU}
\affil[2]{Ordinal Research Institute, Wilmington, DE 19803, USA}
\affil[3]{University of Zagreb, Faculty of Food Technology and Biotechnology, \newline Pierottijeva 6, HR-10000 Zagreb, Croatia, EU}

\begin{document}
\maketitle

\begin{abstract}
In this paper, we present a statistical model of spacetime trajectories based on a finite collection of paths organized into a branched manifold. For each configuration of the branched manifold, we define a Shannon entropy. Given the variational nature of both the action in physics and the entropy in statistical mechanics, we explore the hypothesis that the classical action is proportional to this entropy. Under this assumption, we derive a Wick-rotated version of the path integral that remains finite and exhibits both quantum interference at the microscopic level and classical determinism at the macroscopic scale. In effect, this version of the path integral differs from the standard one because it assigns weights of non-uniform magnitude to different paths. The model suggests that wave function collapse can be interpreted as a consequence of entropy maximization. Although still idealized, this framework provides a possible route toward unifying quantum and classical descriptions within a common finite-entropy structure.
\end{abstract} \hspace{10pt}

\maketitle

\section{Introduction}\label{sec:Introduction}

Dynamical evolution in physics takes different forms depending on the theoretical framework employed. In classical mechanics, systems are typically described by trajectories in state space that evolve deterministically over time. Within such frameworks, including Newtonian dynamics and general relativity, initial conditions uniquely determine future states, and the evolution is governed by well-defined differential equations.

Stochastic models introduce probabilistic elements into the evolution \cite{stochastic_prob}. Here, transitions between states occur with specified probabilities, reflecting either intrinsic randomness or the influence of untracked variables. Stochastic models are frequently used in statistical mechanics and other fields where deterministic descriptions are either impractical or insufficient.

Quantum mechanics offers a different structure altogether. The state of a system is described by a complex-valued amplitude in a Hilbert space, evolving deterministically under a unitary operator that is typically governed by the Schrödinger equation. Despite this deterministic evolution, measurements yield outcomes according to a probability distribution given by the squared magnitude of the amplitude. This combination of continuous evolution and discrete, probabilistic measurement outcomes is a defining feature of quantum theory. Within standard interpretations such as the Copenhagen view, this is accounted for by the postulate of wave function collapse, which assigns probabilities to measurement outcomes via the Born rule.

The path integral formulation provides a useful and widely applicable representation of quantum evolution. In this approach, the transition amplitude between initial and final states is expressed as a sum over all possible paths, each weighted by a complex exponential of the classical action. This formalism, introduced by Feynman and foreshadowed by Dirac, captures quantum interference effects naturally and has proven effective in both conceptual and computational contexts~\cite{Feynman:100771, Dirac1933}. Analogies have been drawn between this approach and the continuum descriptions in statistical or fluid systems, where macroscopic order emerges from an underlying multiplicity of microscopic configurations~\cite{Bohr1949-BOHDWE}.

In this paper, we develop a modification of the path integral framework based on a finite set of paths arranged in a branched manifold structure. Rather than integrating over a continuous space of trajectories, we consider a discrete ensemble of paths with defined branching and intersection points. Each path contributes an amplitude, and the total transition amplitude is given by a weighted sum over the ensemble.

A key aspect of the model is the probabilistic sampling of paths. The sampling distribution is influenced by the structure of the branched manifold, in particular by the frequency of intersections between paths. Configurations with more frequent intersections are assigned higher probabilities, introducing an effective entropic bias toward path cohesion. This results in quantum-like interference at small scales and convergence toward classical behavior at larger scales, where the ensemble becomes dominated by a narrow set of similar trajectories.

The branched manifold is formally represented using a simplicial complex, and the model incorporates a conserved branch weight to ensure consistency across the ensemble. The resulting formulation produces a Wick-rotated version of the path integral with finite amplitudes and a clear mechanism for the transition from quantum to classical behavior. The branched manifold structure is based on the theory of foliations and expanding attractors~\cite{calegari2007foliations, williams1974expanding}. We construct this framework in detail and examine its implications for modeling dynamical systems across different regimes.

\section{Geometric framework: Finite branches}\label{sec:Geometric}

We develop our model within the geometric setting of branched manifolds, which generalize smooth manifolds by permitting singularities at points of branching or intersection. These structures provide the necessary scaffolding to capture the superpositional aspects of quantum theory while maintaining a finite, combinatorially tractable basis for spacetime histories.

The tools introduced in this section form the foundation for subsequent constructions. In particular, we define a simplicial decomposition of the branched manifold and introduce a conserved branch weight. These elements allow for a natural expression of the path structure, branch intersections, and ultimately the entropy associated with each configuration. The computation of this entropy—first at the kinematical level in Section~\ref{sec:Kinematics} and later dynamically in Section~\ref{sec:Dynamics}—relies on the framework developed here.

\subsection{Simplicial decomposition}\label{sec:Simplicial}
An embedded branched $n$-manifold is modeled as an $n$-complex in a higher-dimensional ambient space such that each point admits a well-defined $n$-dimensional tangent space. 
Intuitively, a branched manifold represents multiple overlapping smooth sheets of spacetime that meet or diverge along lower-dimensional junctions. 
This structure allows distinct trajectories to intersect, providing a geometric model for quantum superposition.
To encode quantum superposition geometrically, we posit that spacetime itself is a branched manifold $M$, composed of a finite union of smooth branches~\cite{calegari2007foliations, williams1974expanding}. Each branch carries its own differential structure, ensuring the existence of a local tangent space at every point in $M$.

For concreteness, we work within a Minkowski space of $(n+1)$ dimensional coordinates $(t, x_1, \dots, x_n)$, collectively denoted by $\mathbf{x}$. The physical fields are defined over a fiber bundle $\mathbb{R}^{n+1} \times \Phi$, where $\Phi$ is a finite-dimensional complex vector space encoding internal degrees of freedom~\cite{tu2017differential, lee2013introduction}. A branch $b$ is represented by a smooth section $\varphi: \mathbb{R}^{n+1} \to \Phi$, assigning to each point $\mathbf{x}$ a unique field value $\varphi(\mathbf{x})$. A \emph{path} is then the restriction of a branch to a subset $U \subset \mathbb{R}^{n+1}$, naturally capturing the localized structure of field superposition, which will be discussed in Section \ref{sec:Entropy}.

Furthermore, we introduce a coordinate map $X : M \to \mathbb{R}^{n+1}$ associating to each point in $M$ its spacetime coordinates. We define the inverse image of $X$ as
\begin{equation}
    X^{-1}(\mathbf{x}) = \{ y \in M \mid X(y) = \mathbf{x} \},
\end{equation}
which returns the set of all points in $M$ projecting to $\mathbf{x}$.

To regulate branching, we introduce a conserved, positive \emph{branch weight} $w : M \to \mathbb{R}^+$ satisfying
\begin{equation}
    w(y) \geq L > 0,\quad \forall y \in M,
\end{equation}
where $L$ is constant. Consequently, only a finite number of branches can intersect at any given point. 

To ensure well-posedness of our framework, we assume that spacetime is composed of a large but finite number of branches. This restriction ensures the finiteness of sums appearing in the path integral and entropy expressions. It also reflects the hypothesis that in any physical process, only a finite number of distinct histories are realized. Not only does this discretization avoid divergences, but it is also conceptually aligned with a minimal resolution of spacetime at the fundamental level.

We represent the branched manifold $M$ using a simplicial complex~\cite{goerss1999simplicial, hatcher2002algebraic, regge1961general, ambjorn1998nonperturbative}, where each $(n+1)$-simplex $\sigma$ denotes a simply connected region of spacetime and each $n$-simplex $\tau$ represents a shared face between neighboring simplices. We will consider the case where $w$ is locally constant on each $(n+1)$-simplex $\sigma$ and denote it by $w_\sigma$. The branch weight $w_\sigma$ should be understood as an \emph{intensive} quantity—a local density characterizing the relative occupancy of a branch within each simplex—rather than an extensive total over that region.  

\begin{figure}[H]
         \centering
         \includegraphics[height=4cm]{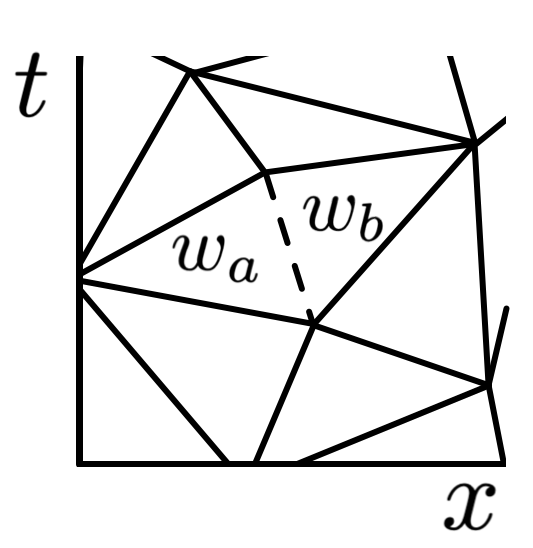}
         \caption{We depict a 1+1-spacetime with just one branch, which is decomposed into several simplices. The value of the field $\varphi$ is not shown in this diagram. The dashed line represents where two simplices meet. The left simplex has branch weight $w_a$, and the right simplex has branch weight $w_b$. The branch weight conservation constraint implies that $w_a=w_b$ in this case.}
        \label{fig:1+1simplices}
\end{figure}

Branch weight conservation across shared boundaries is encoded using the boundary operator $\partial_{n+1}$ from simplicial homology. For an oriented $k$-simplex $\sigma = (v_0, \dots, v_k)$, where $v_i$ are the vertices within the simplex, we define
\begin{equation}
    \partial_k(\sigma) = \sum_{i=0}^k (-1)^i (v_0, \dots, \widehat{v}_i, \dots, v_k),
\end{equation}
where $\widehat{v}_i$ indicates omission of $v_i$. Then we can express the global conservation condition as
\begin{equation}\label{eqn:BranchWeightConservation}
    \sum_{\sigma \in M} w_\sigma\, \partial_{n+1}(\sigma) = 0,
\end{equation}
which illustrates the conservation of branch weights (within a simplicial representation) at each shared $n$-simplex. This structure is illustrated in Figure~\ref{fig:1+1simplices}, where two adjacent $(1+1)$-simplices share a boundary. The boundary operator assigns opposite signs to the shared edge, enforcing the constraint ${w_a = w_b}$ under the assumption of a single branch.

We now turn to the representation of paths. Each path corresponds to a union of simplices, and we define an incidence matrix $A_{\sigma i}$ indicating whether simplex $\sigma$ belongs to path $p_i$:
\begin{equation}
    A_{\sigma i} = 
    \begin{cases}
        1 & \text{if } \sigma \in p_i, \\
        0 & \text{otherwise}.
    \end{cases}
\end{equation}
Branch weights on simplices are then expressed as linear combinations of path weights:
\begin{equation}\label{eqn:BranchWeightBasisChange}
    w_\sigma = \sum_i A_{\sigma i} w_i.
\end{equation}
Refinement of the simplices decomposes each simplex into a set of simplices, $\sigma \rightarrow \{\sigma^{(r)}\}$. 
This decomposition changes the incidence matrix $A_{\sigma i} \rightarrow A^{(r)}_{\sigma^{(r)} i}$, by row replication.
The branch weights $w_\sigma$ are intensive quantities, and therefore the branch weights of the refined simplices are unchanged, $w_{\sigma^{(r)}} = w_\sigma$.

\begin{figure}[H]
    \centering
    \includegraphics[height=4cm]{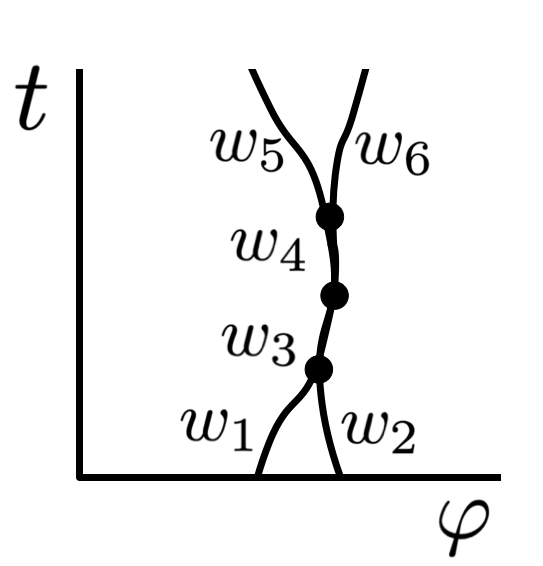}
    \caption{We show a branched 0+1-dimensional spacetime manifold in which branches intersect and subsequently diverge. The 1-simplices along these branches carry branch weights $w_1$ through $w_6$. The 1-simplices intersect at 0-simplices (i.e., vertices).}
    \label{fig:WeightedBranches}
\end{figure}

Figure \ref{fig:WeightedBranches} illustrates a representative branching diagram.
The paths in this figure are denoted by $LML$, $LMR$, $RML$, and $RMR$ (corresponding to left-middle-left, left-middle-right, right-middle-left, and right-middle-right).
Using those paths, we write Equation~(\ref{eqn:BranchWeightBasisChange}) as
\begin{equation}
   \begin{bmatrix}
w_1 \\
w_2 \\
w_3 \\
w_4 \\
w_5 \\
w_6 \\
\end{bmatrix}=
    \begin{bmatrix}
1 & 1 & 0 & 0 \\
0 & 0 & 1 & 1 \\
1 & 1 & 1 & 1 \\
1 & 1 & 1 & 1 \\
1 & 0 & 1 & 0 \\
0 & 1 & 0 & 1 \\
\end{bmatrix}
    \begin{bmatrix}
w_{LML} \\
w_{LMR} \\
w_{RML} \\
w_{RMR}\\
\end{bmatrix}
.
\end{equation}

As we discuss in Section~\ref{sec:Kinematics}, these path weights determine the statistical distribution over path ensembles, linking the simplicial geometry of the branched manifold to entropy-based probabilistic dynamics.

%%%%%%%%%%%%%%%%%%%%%%%%%%%%%%
\subsection{Quantum superposition from finite branches}\label{sec:Superposition}
%%%%%%%%%%%%%%%%%%%%%%%%%%%%%%

We now illustrate how superposition arises in this framework.

Consider two branches $b_1$ and $b_2$ with corresponding fields $\varphi_1$ and $\varphi_2$. At a spacetime point $\mathbf{x}$, these may yield distinct field values, contributing to a superposition. If $\varphi_1(\mathbf{x}) = \varphi_2(\mathbf{x})$, the branches intersect at $\mathbf{x}$, and the resulting superposition effectively reduces to a single state. In general, the number of distinct field values present at $\mathbf{x}$ corresponds to the number of branches with inequivalent values at that point.

Suppose that the manifold $M$ is composed of multiple branches, each equipped with a distinct field $\varphi_i$. At a given spacetime point $\mathbf{x}$, the wave function $\psi(\mathbf{x})$ is defined as the weighted sum over fields on branches:
\begin{equation}\label{eqn:psi}
    \psi(\mathbf{x}) = \sum_{y \in X^{-1}(\mathbf{x})} w(y)\, \varphi(y).
\end{equation}
This sum has a finite number of terms due to the lower bound on $w(y)$ and the branch weight conservation condition. 
Summing over all branches at a point $\mathbf{x}$ gives a total branch weight
\begin{align}\label{eqn:TotalBranchWeight}
    w_T(\mathbf{x}) &= \sum_{y \in X^{-1}(\mathbf{x})} w(y)\,.
\end{align}
The conservation condition in Equation~(\ref{eqn:BranchWeightConservation}) implies that the total branch weight is constant as a function of the coordinates:
\begin{equation}
    \partial_\mu w_T(\mathbf{x}) = 0\,.
\end{equation}

We assume that branches intersect at $\mathbf{x}$ if their associated fields are equivalent as rays, i.e., $\varphi_1(\mathbf{x}) = z\, \varphi_2(\mathbf{x})$ for some nonzero $z \in \mathbb{C}$. These intersections are the quantum interference of this framework. 

Note that $\psi = w \varphi$ is invariant under the rescaling $w \to \alpha w$, $\varphi \to \varphi/\alpha$, highlighting a degeneracy in the representation. Since $w$ is conserved and $\varphi$ is not, such transformations do not yield equivalent systems. Furthermore, if the number of branches exceeds $\dim \Phi$, the decomposition of $\psi$ into field and weight components becomes non-unique. In this framework, the wave function $\psi$ does not encode the entirety of the quantum information; rather, the branched manifold itself, including the distribution of branch weights, serves as the full information carrier. As we will see, this internal structure plays a crucial role in determining entropy and dynamical evolution.

\section{Kinematics: Entropy from conservation of branch weight}\label{sec:Kinematics}

\subsection{Conservation law of branch weight}\label{sec:Conservation}

We now examine how the conservation of branch weight constrains the structure of the branched manifold and affects its entropy. 

Figure~\ref{fig:WeightedBranches} depicts a simple 0+1-dimensional branched manifold with intersecting and separating branches.
Each 1-simplex $\sigma$ is bounded by two 0-simplices $\tau$, and the boundary operator $\partial_{n+1}$ assigns a sign to each vertex based on orientation. Applying the conservation condition from Equation~\eqref{eqn:BranchWeightConservation}, we obtain the following constraints for the configuration in Figure~\ref{fig:WeightedBranches}:
\begin{align}
    w_1 + w_2 - w_3 &= 0, \\
    w_3 - w_4 &= 0, \\
    w_4 - (w_5 + w_6) &= 0.
\end{align}
These relations can be compactly expressed as a linear system:
\begin{equation}\label{eqn:BranchWeightMatrixExample}
    \begin{bmatrix}
        1 & 1 & -1 & 0 & 0 & 0 \\
        0 & 0 &  1 & -1 & 0 & 0 \\
        0 & 0 &  0 &  1 & -1 & -1 \\
    \end{bmatrix}
    \begin{bmatrix}
        w_1 \\ w_2 \\ w_3 \\ w_4 \\ w_5 \\ w_6
    \end{bmatrix} = 0.
\end{equation}
The space of valid branch weight values $w_\sigma$ lies in the null space of this matrix. If, for instance, $w_1$ and $w_2$ are fixed, then the only remaining degree of freedom is the allocation between $w_5$ and $w_6$. 

\begin{figure}[H]
     \centering
     \begin{subfigure}[b]{4cm}
         \centering
         \includegraphics[height=4cm]{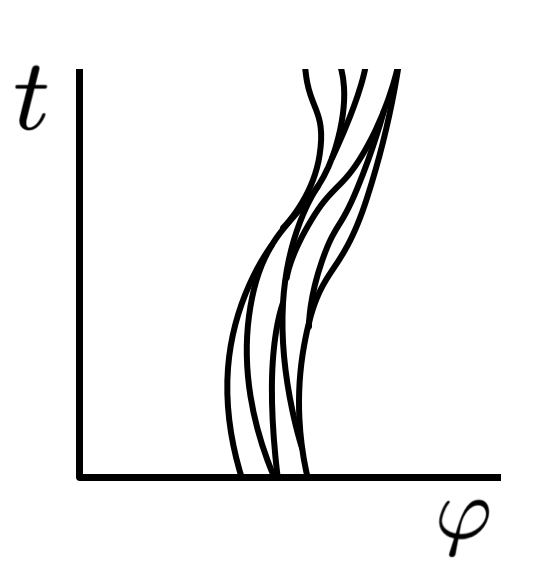}
         \caption{}
     \end{subfigure}
     \hspace{.5cm}
     \begin{subfigure}[b]{4cm}
         \centering
         \includegraphics[height=4cm]{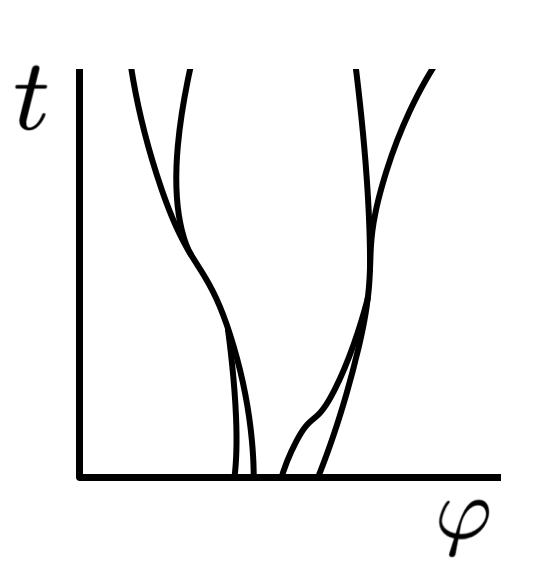}
         \caption{}
     \end{subfigure}
        \caption{ We see two examples of 0+1-dimensional branched manifolds. 
        (\textbf{a}) A branched manifold with high branch cohesion enables frequent intersections. This case has a higher-dimensional null space of branch weights, corresponding to greater entropy. (\textbf{b}) A branched manifold with little cohesion has infrequent intersections. This case has less entropy.}
        \label{fig:BranchCohesionExamples}
\end{figure}

To further illustrate the relationship between branch intersections and degrees of freedom, we compare two examples in Figure~\ref{fig:BranchCohesionExamples}. In Figure~\ref{fig:BranchCohesionExamples}(a), frequent intersections among branches lead to a high-dimensional null space for branch weights. In Figure~\ref{fig:BranchCohesionExamples}(b), branches have fewer intersections and a correspondingly lower-dimensional null space.

In the simplicial decomposition of the branched manifold, the branch weight conservation constraint is given by Equation~(\ref{eqn:BranchWeightConservation}).
We can trivially convert the boundary operator $\partial_{n+1}(\sigma)$ to a matrix: 
\begin{equation}
    D_{\tau\sigma} =
    \begin{cases}
        +1, & \text{if } \tau \text{ appears positively in } \partial_{n+1}(\sigma), \\
        -1, & \text{if } \tau \text{ appears negatively in } \partial_{n+1}(\sigma), \\
        \,\,0, & \text{otherwise}.
    \end{cases}
\end{equation}
The conservation law then takes the form
\begin{equation}\label{eqn:DTauSigma}
    \sum_{\sigma \in M} D_{\tau\sigma} w_\sigma = 0,
\end{equation}
which asserts that the vector of branch weights $w_\sigma$ lies in the null space of $D_{\tau\sigma}$. Valid configurations must also satisfy the lower bound $w_\sigma \ge L$ for the fixed constant $L > 0$.

The dimensionality of the null space of $D_{\tau\sigma}$ determines the number of independent degrees of freedom in the branch weight configuration. As the number of adjacent $(n+1)$-simplices sharing a common $n$-simplex increases, so too does the dimensionality of the null space. This implies that more frequent branch intersections enlarge the space of admissible branch weight configurations, thereby increasing the entropy associated with the manifold.

While the branch weight entropy captures correlations between intersecting branches, it does not capture the full statistical behavior of individual paths. To analyze the latter, it is useful to reparameterize branch weights in terms of paths, using a change of basis from simplices to paths. As we shall see, the entropy of paths plays a critical role in the probabilistic formulation of dynamics.

\subsection{Probability space of branched manifolds}

The probability space of branched manifolds is 
\begin{equation}
    \Omega_M = \left\{\, (M, w, \varphi)
    \right\},
\end{equation}
where each element specifies a branched manifold $M$, a branch weight configuration $w$, and a field assignment $\varphi$. 
From Equation~(\ref{eqn:DTauSigma}), we see that the branch weight is conserved across branch intersections, and this implies that there is a fixed total branch weight $w_T$ over every point. 
Because of the discrete properties of branch intersections, these constraints are expressed most cleanly on a simplicial decomposition of $M$.
However, there are multiple ways to decompose any branched manifold into simplices. 
Let $\sigma(\textbf{x})$ be the set of all simplices $\sigma \in M$ that are over a point $\textbf{x}\in \mathbb{R}^{n+1}$. 
Then the constraint of constant total branch weight is
\begin{equation}
    \sum_{\sigma\in \sigma(\textbf{x})} w_\sigma = w_T.
\end{equation}
We now describe the probability space of branched manifolds, assuming that each branched manifold $M$ is decomposed into simplices $\sigma$ with corresponding branch weights $w_\sigma$,
\begin{equation}
    \Omega_M = \left\{\, \chi = (M, w, \varphi) \;\middle|\;
    \sum_{\sigma \in M} D_{\tau\sigma} w_\sigma = 0,\;
    \sum_{\sigma\in \sigma(\textbf{x})} w_\sigma = w_T,\;
    w_\sigma \ge L > 0
    \right\}.
\end{equation}
This space is subject to the constraint that any two elements of this space are equivalent if their values of $w$ and  $\varphi$ are the same over every point of $\mathbb{R}^{n+1}$.

Calculation of probability in this space is a challenging combinatorics problem. 
Some of the computational difficulty may be alleviated by alternative descriptions of the space.
However, the probability space describes generic physical systems and a formula for probability or entropy would also constitute a formula for the prediction of any physical system. 
Such a formula is uncomputable~\cite{moore1990unpredictability}.
As with the theory of computation, the general problem does not have a solution, but there may be opportunities to explore special cases and approximate their likely outcomes.

\subsection{Entropy of the branched manifold}\label{sec:Entropy}

The branched manifold exhibits two distinct sources of randomness: the stochastic fluctuations of the field $\varphi$ and the freedom in assigning branch weights $w$. The interplay between these degrees of freedom determines the equilibrium behavior of the system. In particular, a dynamic balance emerges between the entropic tendency of fields to differentiate (favoring branch separation) and the entropy of branch weights (favoring intersection and cohesion).

To formalize this, we define a map $\Psi$ that assigns to each branched manifold $M$ a corresponding wave function $\psi_M$. At a point $\mathbf{x} \in \mathbb{R}^{n+1}$, the wave function is given by $\psi_M(\mathbf{x})$, as defined in Equation~(\ref{eqn:psi}). 
Thus, $\psi_M(\mathbf{x}) \in \Phi$ lies in the space of field values, and $\psi_M$ itself defines a smooth section of the fiber bundle $\mathbb{R}^{n+1} \times \Phi$. 
In this sense, the wave function $\psi_M$ can be interpreted as an effective, coarse-grained branch.

Restricting to an open subset $ U \subset \mathbb{R}^{n+1} $, we may identify the restriction of $\psi_M$ with a path $ p = \psi_M|_U $. To quantify the number of distinct branched manifolds that are consistent with a given path, we define the inverse image of this map as
\begin{equation}
    \Psi^{-1}(p) = \{ \chi \mid \Psi(\chi) = p \},
\end{equation}
where we implicitly assume that the total branch weight on each branched manifold $\chi$ is the fixed total branch weight $w_T$. The set $ \Psi^{-1}(p) $ thus consists of all branched manifolds that yield the same effective path $ p $. In this sense and in analogy with statistical mechanics, the elements $ \chi \in \Psi^{-1}(p) $ represent the microstates corresponding to the path $ p $.

The randomness in $\varphi$ and $w$ induces a probability measure $P(\chi)$ on the space of branched manifolds. We use this to define a Shannon entropy over the ensemble of microstates consistent with the path $ p $:
\begin{equation}
    S_{\mathrm{en}}[p] = -\int_{\Psi^{-1}(p)} P(\chi)\, \ln P(\chi)\, \mathrm{d}\chi.
\end{equation}
Unlike conventional thermodynamic entropy, which is typically defined at a fixed time, the entropy $ S_{\mathrm{en}}[p] $ is defined over both space and time, reflecting the full kinematical history encoded by the path $ p $. The entropy $ S_{\mathrm{en}}[p] $ quantifies the statistical weight of a given coarse-grained trajectory in the branched manifold and will serve as a central quantity in the dynamical considerations to follow.

\section{Dynamics: Entropic action principle}\label{sec:Dynamics}

If the branched manifold is in a state of local equilibrium, then small variations in the path $p$ do not increase the entropy. In this regime, the entropy functional is stationary:
\begin{equation}
    \delta S_{\mathrm{en}}[p] = 0.
\end{equation}
This observation motivates the use of entropy as a variational principle. Other work has used the idea of relating action to entropy \cite{Mathur2021, Hamada2024}. In the discussion that follows, we propose that the dynamics of the system can be captured through an action proportional to $S_{\mathrm{en}}$, the Shannon entropy on branched spacetime manifolds. 

We define the action functional $S[p]$ as a linear rescaling of the entropy:
\begin{equation}
\label{eq:entropy_action}
    S[p] = -\alpha S_{\mathrm{en}}[p],
\end{equation}
where $\alpha > 0$ is a constant that converts entropy into units of action. The negative sign ensures that paths with higher entropy correspond to lower action, consistent with the usual extremization principle.

While the field $\varphi$ exhibits a stochastic behavior governed by an underlying probability distribution, the precise nature of this distribution is not fully specified by our assumptions. To proceed, we postulate that the induced dynamics for the wave function $\psi$ are consistent with the conventional unitary evolution of quantum mechanics. 

\subsection{The path integral as a linear approximation of a nonlinear model}\label{sec:non-linear_models}

The finite lower bound $w \geq L$ on branch weights introduces a nonlinearity into the model. In particular, a branched manifold with total branch weight $w_T = L$ consists of a single branch, but if $w_T < L$, no valid branching configuration exists. This discontinuity implies that the path ensemble cannot be scaled arbitrarily and reflects the underlying discreteness of the system.

Such nonlinearities are crucial for modeling quantum measurement, as we explore in Sections~\ref{sec:Measurement} and~\ref{sec:Discussion}. They represent departures from the standard linear formalism of quantum theory, which is otherwise recovered as an effective approximation in regimes where the branching structure is sufficiently dense and near equilibrium.

Moreover, a quantum theory is fundamentally linear and unitary, yet measurement introduces abrupt, nonlinear updates to the wave function~\cite{Ghirardi1986, Bassi2013, Hu2023}. A model that seeks to incorporate both regimes must account for this dichotomy: the deterministic linear evolution of isolated systems and the nonlinear, stochastic behavior during measurement. The entropy-based framework proposed here accommodates both by treating linear quantum mechanics as a limiting case \cite{caticha2017entropicdynamicsquantummechanics}.

We now derive an approximate path integral based on the entropic action $S[p]$.  
Let $M_{c_I c_F}$ denote the set of paths beginning in configuration $c_I$ at time $t_I$ and ending in $c_F$ at time $t_F$. 
Applying the standard time-slicing argument~\cite{Feynman:100771, Dirac1933}, we associate a phase $e^{(i/\hbar) S[p]}$ to each path $p$.
Each path $p_i \in M_{c_I c_F}$ is assigned a weight $w_i$, as defined in Section~\ref{sec:Simplicial}.
From Equation~(\ref{eqn:psi}), the wave function is a weighted sum over the branches. 
In the transition from the initial configuration $c_I$ to the final configuration $c_F$, each path contributes a weighted phase $w_i\, e^{(i/\hbar) S[p_i]}$.
Summing the contributions of each path in $M_{c_I c_F}$ gives
\begin{equation}\label{eqn:SumOfPaths}
    \tilde{Z} = \sum_{p_i \in M_{c_I c_F}} w_i\, e^{(i/\hbar) S[p_i]}.
\end{equation}

Since the complete branching structure is typically unknown, direct computation of $\tilde{Z}$ is impractical. However, we can calculate its expected value $E[\tilde{Z}]$ 
by integrating $\tilde{Z}$ over all branched manifolds that have total branch weight $w_T$.
For a path $p_i$, its branch weight $w_i$ depends on the branched manifold of which $p_i$ is a part, and we use statistical properties of the branch weight to determine its contribution to $E[\tilde{Z}]$.

In the linear limit of quantum mechanics, the paths of a branched manifold are independent of each other. 
The branch weight of a particular path is a degree of freedom. 
For independent paths and an entropy maximizing distribution of branch weights, a branch weight has the same probability distribution regardless of the path to which it is associated.
Every branch weight therefore has the same expected value, $w_E$.
If a path is not present in a given branched manifold, then its corresponding branch weight is effectively zero. 
We write $\mathbb{P}(p_i)$ to indicate the probability that a path $p_i$ is in a given branched manifold. 
A path $p_i$ is either present in a branched manifold and has a branch weight with expected value $w_E$, or else it is not present and its branch weight is zero.
Therefore, for a particular path $p_i$, the expected value of its branch weight $w_i$ over all branched manifolds is
\begin{equation}
    E[w_i] = w_E \mathbb{P}(p_i).
\end{equation}
As with simplex branch weights $w_\sigma$, the per-path branch weight $w_i$ is likewise an \emph{intensive} property. Its normalization is fixed by the conserved total branch weight $w_T$.

As established in Section~\ref{sec:Entropy}, the entropy $S_{\mathrm{en}}[p]$ is the logarithm of the number of branched manifolds that yield the wave function $p$. The probability of a particular path $p_i$ is proportional to the exponential of its entropy. Using the action-entropy relation, this implies a probability
\begin{equation}
    \mathbb{P}(p_i) \propto e^{-k S[p_i]}
\end{equation}
for some constant $k > 0$. 

The action of a path $S[p_i]$ depends only on the path $p_i$.
The factor $e^{(i/\hbar) S[p_i]}$ is therefore constant when we integrate over all branched manifolds to calculate the expectation value 
\begin{equation}
    E\left[ w_i\, e^{(i/\hbar) S[p_i]} \right] = E\left[ w_i\right]\, e^{(i/\hbar) S[p_i]} =w_E \mathbb{P}(p_i) e^{(i/\hbar) S[p_i]} \propto  e^{(i/\hbar - k) S[p_i]}.
    \end{equation}
Neglecting normalization constants and converting the sum to an integral over paths, we obtain
\begin{equation}\label{eqn:E[Z]}
    E[\tilde{Z}] = \zeta \int e^{(i/\hbar - k) S[p]}\, \mathcal{D}p,
\end{equation}
where $\mathcal{D}p$ is the path measure and $\zeta$ is a normalization factor. This is a Wick-rotated variant of the standard Feynman path integral~\cite{wick1954properties}, where due to Equation~\eqref{eq:entropy_action} the entropy acts as an effective action in Equation~\eqref{eqn:E[Z]}.

In the standard Feynman path integral formulation, all kinematically allowed paths contribute equally in magnitude—each path carries the same amplitude weight, differing only in phase through the classical action. 
However, in the present framework based on branched manifolds and entropic dynamics, not all paths are treated equally. 
Instead, each path carries a distinct probability, reflecting its statistical likelihood of occurring within the ensemble of possible evolutions. 
Physically, this means that a path is more likely to appear if it is associated with higher entropy—that is, if it corresponds to a greater number of compatible micro-configurations within the underlying simplicial complex. 
The weighting encodes an entropic preference, biasing the system toward histories that are more combinatorially accessible, or more probable in an informational sense. 
This breaks the uniform weighting assumption of standard quantum mechanics and introduces an effective measure over paths.

The exponential factor $e^{-k S[p]}$ in Equation~\eqref{eqn:E[Z]} describes the
\emph{statistical weighting} of individual paths within the ensemble prior to
coherent summation. Paths with large action values correspond to lower
entropy and therefore carry exponentially smaller probabilities of occurrence.
It is important to note that this suppression acts on the sampling probability
of paths, not directly on the magnitude of the final interference amplitude
$|\tilde Z|$. Once the ensemble of admissible paths has been drawn according
to these probabilities, their complex phases $e^{iS[p]/\hbar}$ still interfere
coherently in the summation. In this way, high--action paths are
statistically rare, but their contributions remain subject to interference
with other paths in forming the total amplitude. This distinction clarifies
that the model’s outlier control mechanism is fundamentally probabilistic,
arising from entropy--weighted sampling rather than post--summation damping.

Equation~(\ref{eqn:E[Z]}) provides a natural context for interpreting the Wick rotation, the transformation to imaginary time used in many path integral treatments. In traditional quantum mechanics, the Wick rotation serves to regularize divergences, ensure convergence of the integral, and define a correspondence with statistical mechanics \cite{wick1954properties, Schlingemann1999}. In our setting, the Wick rotation can also be viewed as selecting the dominant entropic configurations, allowing entropy-weighted paths to contribute with real-valued probabilities, rather than oscillatory phases. This not only improves mathematical control but also clarifies the physical interpretation of the path sum as a probability-weighted statistical ensemble.

For paths far from equilibrium, the nonlinear constraints of the model render this expectation inaccurate. Such nonlinear effects are central to the behavior of measurements, which we now examine.

\subsection{Collapse}\label{sec:Measurement}

The entropy-weighted path integral in Equation~\eqref{eqn:E[Z]} favors configurations of high entropy and low action. However, this condition alone does not enforce wave function collapse. For example, a measurement process may allow several macroscopically distinct outcomes, each associated with comparable entropy and action. In such cases, multiple coarse-grained paths may remain dynamically viable.

The collapse mechanism arises instead from the constraint on branch cohesion encoded in Equation~\eqref{eqn:DTauSigma}. 
Entropy is maximized when branch intersections are frequent, implying that nearby branches must remain similar. 
A superposition of macroscopically distinct states (e.g., different measurement outcomes) disrupts this cohesion. Pathological field configurations between intersecting branches would incur large action penalties and are, therefore, statistically suppressed.

Thus, to preserve high entropy and low action, the branched manifold selects a configuration in which all branches align with a single measurement outcome. This alignment constitutes the wave function collapse in our framework.

\begin{figure}[H]
    \centering
    \includegraphics[height=4cm]{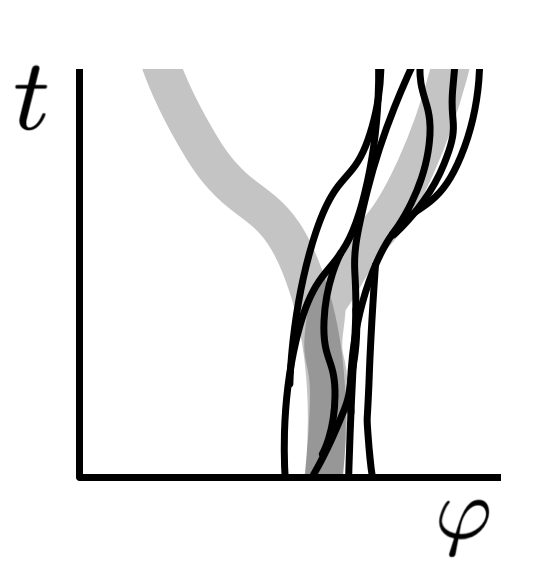}
    \caption{Gray lines indicate possible field configurations corresponding to two distinct measurement outcomes. After the measurement event, the field values diverge significantly. Thin lines represent branches of the spacetime manifold. The entropy constraint requires that branches remain close to each other. Since they cannot simultaneously remain close to both outcomes, the manifold collapses to a single result (here, the right-hand outcome) to maximize entropy.}
    \label{fig:Measurement}
\end{figure}

\subsection{Entropic collapse condition}

In the finite-entropy framework, wave function collapse corresponds to the
system selecting a single macroscopic branch when entropic pressure prevents
the coexistence of distinct configurations.  We now formulate this quantitatively.

For a system with two competing subsystems that have total branch weights $w_1$ and $w_2$, the entropy can be approximated as a function of the subsystem branch weights, $S_{en}(w_1,w_2)$. 
If $\partial S_{en}/\partial w_i \ne 0$, then the system will adjust branch weights to restore equilibrium. 
If the adjustment process remains out of equilibrium until either $w_1=0$ or $w_2=0$, then the system collapses.

Assume an uncollapsed region of spacetime: a subset $U\subset \mathbb{R}^{n+1}$ such that the paths $p$ over $U$ can be partitioned into disjoint subsets of non-intersecting paths.   
The path partition implies that the n+1-simplices $\sigma$ can likewise be partitioned, and this further implies that any n-simplex $\tau$ can only connect n+1-simplices $\sigma$ that are within the same partition.
This implies a block decomposition $D_{\tau \sigma} = \oplus_j D^j_{\tau\sigma}$ with each $D^j$ corresponding to one of the subsets of the partition.
If $D_{\tau\sigma}$ can be block decomposed, then it has a lower-dimensional null space than a generic $D_{\tau\sigma}$ and therefore lower branch weight entropy.
The dimension of $D_{\tau\sigma}$ scales linearly in the number of simplices over $U$, which is linear in the volume of $U$. The entropy deficit for the block decomposition therefore grows linearly in the volume of the spacetime region $U$. 
This implies a threshold size for a superposition beyond which collapse is effectively inevitable.

\subsection{A 0+1-dimensional sample model}

To illustrate how the entropic collapse condition operates in the simplest setting, we consider a 0+1-dimensional toy model that is the union of two branches $b_1$ and $b_2$ that have fields $\varphi_1$ and $\varphi_2$, respectively.
We will consider a collapsed and an uncollapsed case.

For the uncollapsed case, $M_A$, the branches $b_1$ and $b_2$ never intersect.
This implies the branch weights $w_1$ and $w_2$ are constant and the entropy in branch weight allocation is zero.
The entropy of $M_{A}$ is due only to the random variations of the fields $\varphi_1$ and $\varphi_2$.
That entropy is linear over time duration $T$.
For each branch, the entropy is equal to $bT$, for some constant $b$. 
The total entropy is 
\begin{equation}
    S_{en}[M_A] = 2bT.
\end{equation}

For the collapsed case, $M_B$, the branches $b_1$ and $b_2$ intersect at every integer time $t$ and then separate again. 
For total branch weight $w_T$, there is $w_T-2L$ of excess branch weight that is allocated randomly at each branch separation. 
For $T$ time steps, this gives an entropy in branch weight allocation that is $S_{en,w}[M_B] = T\ln(w_T-2L)$.
The field values $\varphi_1$ and $\varphi_2$ must stay similar in order to frequently recombine.
This implies that they have less field entropy than two independent branches, but they have more field entropy than a single branch: ${bT < S_{en,\varphi}[M_B]<2bT}$.
The total entropy is 
\begin{equation}
    S_{en}[M_B] = S_{en,w}[M_B]+S_{en,\varphi}[M_B] = T\ln (w_T-2L)+S_{en,\varphi}[M_B]
\end{equation}
If the total branch weight $w_T > e^b+2L$, then $S_{en}[M_B]>S_{en}[M_A]$ and the collapsed state is entropically favorable.  

Another way to express the entropy in branch weight allocation is by counting the branches.
The collapsed manifold, $M_B$, contains branches which sometimes follow $b_1$ and other times follow $b_2$. 
Let $f:\mathbb{Z}\rightarrow \{1,2\}$ be a function from the integers to the set $\{1,2\}$ and let $b_f$ be the branch which follows branch $b_{f(t)}$ for the time step from $t$ to $t+1$.
If we consider paths from time $t=0$ to time $t=T$, then there are $2^T$ such paths. 
Equation~(\ref{eqn:BranchWeightBasisChange}) describes how branch weight allocation on simplices $\sigma$ of $M_B$ is equivalent to branch weight allocation on these paths $p_i$.

\section{Probability and unitarity}

\subsection{A combination of Markov process and unitary evolution}

The branched manifold can be approximated as a combination of a Markov process and a unitary evolution. 
In this approximation, the branches are a cluster around a single classical path $p_C$ that is a moving Markov center (as depicted in Figure~\ref{fig:Measurement}).
We can take the continuum limit of the behavior of both the cluster and the Markov process.
The probability of a Markov chain is the product of the probabilities of its subchains.
The probability of the classical path is therefore ${P[p_C] \propto e^{-kS[p_C]}}$. 

To describe unitary evolution of the cluster of branches, we approximate the branch weight expectation value using a cutoff that is conditional on branches staying similar to the classical path $p_C$.
Of course, this cannot be done in advance because the classical path is not known until the wave function has collapsed. 
However, after the classical path is known, we can design a conditional expectation value for the branch weight $w_i$ corresponding to a path $p_i$.
\begin{equation}
    E[w_i \,\vert\, p_C] = 
    \begin{cases}
    w_E,& \text{if $p_i$ is similar to $p_C$}\\
    0,              & \text{otherwise}
\end{cases}
\end{equation}
For paths $p_i$ that are similar to $p_C$,
\begin{equation}
    E[w_i e^{(i/\hbar)S[p_i]} \, \vert \, p_C] = E[w_i\,\vert\,p_C]\,e^{(i/\hbar)S[p_i]} = w_E e^{(i/\hbar)S[p_i]}
\end{equation}
For a path integral that integrates only over paths that are similar to the classical path, the result is unitary. 
This path selection reproduces the path integral of quantum field theory with the similarity cutoff serving the role of regularization.
The accuracy of the cutoff approximation depends on how sharply branch weight decreases as paths become less similar to the classical path.

\subsection{The cutoff approximation}

A discontinuous cutoff, as we modeled in the previous section, provides a convenient modification to the expectation value for the branch weight, but a discontinuous cutoff is physically implausible. 
Alternatively, there exist analytic functions that are approximately linear for small deviations $u$ from the classical path, but which saturate for large deviations. A convenient and analytically smooth choice is
\begin{equation}
    f(u) = u_0 \tanh\!\left(\frac{u}{u_0}\right),
\end{equation}
where $u_0$ sets the characteristic scale of the response. For $|u| \ll u_0$,
$f(u) \approx u$, recovering the linear regime used in the weak-coupling
approximation. For $|u| \gg u_0$, $f(u)$ saturates at $\pm u_0$, ensuring that
the effective field and entropy–action coupling remain bounded. This choice
regularizes the strong-coupling limit and prevents unphysical divergences in
the effective action, while retaining analytic continuity with the weak-field
behavior.

To characterize nonlinear behavior independently of normalization effects in
the probability distribution, it is convenient to introduce the log--odds
ratio of path probabilities,
\begin{equation}
    D(u) = \ln\!\left[\frac{P(u)}{1 - P(u)}\right],
\end{equation}
where $P(u)$ is the normalized probability of a path parameterized by its
local deviation $u$ from equilibrium. The nonlinear response is then captured
by the observable
\begin{equation}
    J(u) = \frac{1}{4}\big[D(u) - D(0)\big],
\end{equation}
which serves as the sufficient statistic conjugate to the effective coupling
field, in analogy to how magnetization is conjugate to magnetic field in the
Ising model. Using the bounded response function $f(u) = u_0\tanh(u/u_0)$,
the onset of nonlinearity occurs when $|u|\!\sim\!u_0$, marking the
saturation scale of the system in the $J(u)$ channel. This formulation
provides a quantitative and testable measure of the nonlinear regime that is
consistent with the model’s entropic foundation.
 
\subsection{The Born rule}

To calculate the probability of a measurement result, decompose the wave function into a sum of disjoint results $r$ in a set of possible results $R$.
\begin{equation}
    \psi = \sum_{r\in R} \psi_r
\end{equation}
The probability of a particular result $r$ is proportional to the measure of the configuration space in which the branched manifold transitions from the superposition of multiple measurement results to that single result. 
It is natural then for the wave function collapse to have probability
\begin{equation}
    P = \langle\psi \,\vert\, \psi_r \rangle.
\end{equation}
If the wave function is normalized ($\langle\psi \,\vert\, \psi \rangle = 1$) and the results are orthogonal ($a\ne b \rightarrow \langle\psi_a \,\vert\, \psi_b \rangle = 0$) then the probability of a particular result $r$ follows the Born rule:
\begin{equation}
    P(\psi_r) = \langle\psi \,\vert\, \psi_r \rangle = \langle\psi_r \,\vert\, \psi_r \rangle = \vert \psi_r \vert ^2.
\end{equation}

\subsection{The continuum limit}

The nonlinear features of this branched manifold model arise from the lower bound on the branch weight, $w \ge L >0$.
Indeed, the nonlinear features of the model persist for any value of $L>0$. 
There is a qualitative difference in the behavior for the case $L=0$, with a transition that resembles a phase change. 
In section~\ref{sec:Superposition}, we described a scaling $w\rightarrow \alpha w$ and $\varphi \rightarrow \varphi/\alpha$ that leaves $\psi=w\varphi$ unchanged.
If the lower bound $L=0$, then this scaling has no physical significance.
In this case, a branched manifold with wave function $\psi$ behaves identically regardless of its total branch weight $w_T$.
For that reason, the branch weight in such a system is a redundant feature. 
Such a model then matches the assumptions of quantum field theory and its predictions are the same.

\section{Discussion}\label{sec:Discussion}

The finite path-integral framework replaces the continuum of histories with a finite, entropy-regulated ensemble. This single modification yields convergence, reproduces the quantum–classical transition, and offers a natural interpretation of wave-function collapse.

The atomic model of matter excels at bridging the gap between discrete, microscopic particles and smooth, macroscopic behavior: by averaging over the random motions of a finite collection of atoms near equilibrium, one derives continuum equations such as the heat, wave, and diffusion equations.  
In a parallel vein, Jacobson \cite{jacobson1995thermodynamics} showed that imposing a finite entropy density on quantum fields—and identifying that entropy with the Unruh effect—forces spacetime geometry to obey the Einstein equation, thereby linking quantum thermodynamics to classical gravity.  

Our model adopts the same guiding principle of finite entropy, but implements it through a branched manifold that mimics Feynman’s path integral while summing only over a bounded, discrete set of trajectories. 
To forestall an unmanageable proliferation of histories, each branch carries a conserved, strictly positive branch weight, ensuring a lower bound on its contribution. 
This construction preserves the essential interference structure of quantum mechanics yet guarantees convergence and a well‑defined entropy, uniting microscopic stochasticity with macroscopic differential laws in a single finite‑entropy framework.

The wave function $\psi$ in our model emerges naturally as the projection of the underlying branched manifold onto the usual Hilbert‐space description.  
Just as macroscopic variables like temperature carry an associated entropy, we associate $\psi$ with an entropy functional $S_{en}[\psi]$, whose local maximum characterizes equilibrium and plays the same role as a classical action.  
When the manifold hovers near this maximum‑entropy state, $\psi$ evolves deterministically—recovering unitary Schrödinger dynamics—and only small fluctuations around the peak are allowed.  
The likelihood of each individual branch can be determined by the amount by which that path changes the total entropy of the branched manifold, and summing these weighted contributions over our finite set of paths yields a discrete analogue of the Feynman path integral. 
This finite sum of amplitudes is a discrete version of the quantum path integral. 
However, if a strong interaction, measurement, or environmental coupling pushes the manifold away from one maximum and to another, the condition $ \delta S_{en}[\psi]=0$ no longer holds globally.  
The evolution then becomes intrinsically non‑deterministic—mirroring wave‑function collapse—yet the conserved branch weights exert an entropic cohesion that keeps all branches closely similar, ensuring a well‑defined single macroscopic outcome even though its precise realization remains unpredictable.

Viewed more broadly, quantum superposition in this framework is nothing more than a finite ensemble of branches drawn randomly from all possible branched manifolds.  
Unlike the standard path integral, which integrates over an uncountable infinity of histories, our model requires only a finite number of branches to produce interference, so long as more than one branch contributes appreciably. 
By assuming a finite ensemble of branches, we derived a Wick-rotated path integral that has desirable convergence properties and resembles classical mechanics in the limit of large action. 
Moreover, an intrinsic entropic \textit{pressure} favors frequent intersections among branches, effectively forcing them to recombine and collapse onto a single trajectory when driven far from equilibrium.  
In this way, a single finite‑entropy principle underlies reversible quantum interference at the microscopic scale and irreversible, collapse‑like behavior at the macroscopic scale—bridging the quantum‑to‑classical divide without invoking any external measurement postulates.

Looking forward, it will be important to apply the branched‑manifold construction to concrete quantum systems, such as spin chains or harmonic oscillators, to quantify how the distribution of branch weights affects interference  and convergence toward classical trajectories. 
Extending the framework to mixed states and open systems will clarify how intrinsic entropic collapse competes with, or complements, standard decoherence by an external environment. 
Moreover, exploring how multipartite entanglement and many‑body interactions shape the branched manifold may reveal new scaling laws for the quantum‑to‑classical transition. 
Finally, one might ask whether the branch‑weight entropy principle can guide the design of novel quantum control and error‑correction schemes: by actively steering the manifold toward high‑entropy configurations, we may be able to stabilize coherent superpositions or guide faster collapse when a measurement outcome is required. 
Each of these avenues stands to deepen our understanding of how a single, finite‑entropy principle orchestrates both quantum interference and classical emergence.

%\section{Acknowledgments}

%The authors thank Garrett Biehle, Bassem Sabra, and Sebastian Zaj\k{a}c for their helpful perspectives.

\appendix

\section{Derivation of the entropy-weighted path integral}\label{app:Derivation}

We provide a detailed derivation of Eq.~(\ref{eqn:E[Z]}) and the approximations under
which it holds.

The ensemble-averaged amplitude over branched manifolds is
\begin{equation}
E[\tilde Z] = E\!\left[\sum_i w_i\, e^{iS[p_i]/\hbar}\right]
             = \sum_i E\!\left[w_i\, e^{iS[p_i]/\hbar}\right].
\end{equation}

We assume that for sufficiently large ensembles of branched manifolds:
\begin{enumerate}
    \item The branch weights $\{w_i\}$ are independent random variables
          uncorrelated with the actions $\{S[p_i]\}$;
    \item The ensemble size is large enough that higher-order correlations
          of $w_i$ are negligible (law of large numbers).
\end{enumerate}
Under these assumptions, the expectation factorizes to leading order:
\begin{equation}
E[w_i\, e^{iS[p_i]/\hbar}]
  = E[w_i]\,E[e^{iS[p_i]/\hbar}] + \mathcal{O}(\mathrm{Cov}[w_i,S[p_i]]).
\end{equation}

The expected branch weight is given by
$E[w_i] = w_E\,\mathbb{P}(p_i)$,
where the probability of path $p_i$ in the ensemble follows the entropic
distribution
\begin{equation}
  \mathbb{P}(p_i)
  = \frac{1}{Z_P}\,e^{-kS[p_i]},
\end{equation}
with normalization factor $Z_P = \sum_i e^{-kS[p_i]}$.
Combining these expressions yields
\begin{equation}
E[w_i\, e^{iS[p_i]/\hbar}]
  \propto e^{(i/\hbar - k) S[p_i]}.
\end{equation}

Substituting this into the sum and replacing the discrete ensemble with a
continuous measure $\mathcal{D}p$ gives
\begin{equation}
E[\tilde Z]
  = \zeta \int e^{(i/\hbar - k) S[p]} \, \mathcal{D}p,
\end{equation}
which is Eq.~(\ref{eqn:E[Z]}) in the main text.  The exponential damping term
$e^{-kS[p]}$ originates from the entropic probability measure, while the
factorization step is valid in the large-ensemble (weak-correlation) limit.

If correlations between branch weights and actions are retained, the exact
expectation can be written using a cumulant expansion:
\begin{equation}
E[w_i e^{iS[p_i]/\hbar}]
  = E[w_i]\,e^{iE[S[p_i]]/\hbar}
    \exp\!\left[-\frac{1}{2\hbar^2}
    \mathrm{Var}[S[p_i]] + \ldots \right],
\end{equation}
which yields subleading corrections to the simple damping factor.
These corrections are suppressed in the large-ensemble limit but may
become relevant for small systems or near strong coupling.

\printbibliography %Prints bibliography

\end{document}